\documentclass[aps,prb,preprint,superscriptaddress,showpacs]{revtex4}

\usepackage{graphicx}

\def\ket#1{\mathinner{|{#1}\rangle}}

{\catcode`\|=\active 
  \gdef\Braket#1{\left<\mathcode`\|"8000\let|\BraVert {#1}\right>}}
\def\BraVert{\egroup\,\mid@vertical\,\bgroup}

\begin{document}

\title{High energy photoelectron diffraction:\\
model calculations and future possibilities
}


\author{Aimo Winkelmann}
\email[]{winkelm@mpi-halle.mpg.de}
\affiliation{Max-Planck
Institut f\"{u}r Mikrostrukturphysik, Weinberg 2, D-06120 Halle
(Saale), Germany}

\author{Charles S. Fadley}
\affiliation{Dept. of Physics, University of California Davis, Davis, CA 95616 USA}
\affiliation{Materials Sciences Division, Lawrence Berkeley National Laboratory, Berkeley, CA 94720}

\author{F. Javier Garcia de Abajo}
\affiliation{Instituto de Optica -- CSIC, Serrano 121, 28006 Madrid, Spain}



\date{\today}

\begin{abstract}
We discuss the theoretical modelling of x-ray photoelectron diffraction (XPD) with hard x-ray excitation at up to 20\,keV, using the dynamical theory of electron diffraction to illustrate the characteristic aspects of diffraction patterns resulting from such localized emission sources in a multi-layer crystal. 
 We show via dynamical calculations for diamond, Si, and Fe that the dynamical theory well predicts available current data for lower energies around 1\,keV, and that the patterns for energies above about 1\,keV are dominated by Kikuchi bands which are created by the dynamical scattering of electrons from lattice planes. The origin of the fine structure in such bands is discussed from the point of view of atomic positions in the unit cell.  The profiles and positions of the element-specific photoelectron Kikuchi bands are found to be sensitive to lattice distortions (e.g. a  1\% tetragonal distortion) and the position of impurities or dopants with respect to lattice sites.    We also compare the dynamical calculations to results from a cluster model that is more often used to describe lower-energy XPD.  We conclude that \mbox{hard XPD} (HXPD) should be capable of providing unique bulk-sensitive structural information for a wide variety of complex materials in future experiments.

\end{abstract}




\maketitle


\section{Introduction}

The method of x-ray photoelectron diffraction (XPD) is a powerful tool for the analysis of surface atomic structure, including adsorbates and overlayer growth. By measuring the angular intensity of photoelectrons excited by x-rays and comparing the experimental data with simulations, element-specific information on the surface crystallography of the sample can be gained \cite{fadley:synchrotron}. 
The typical electron kinetic energies in such experiments range from about 100\,eV to 1500\,eV.  Recently, however, an increasing number of photoemission studies have been aimed at developing and applying hard x-ray photoelectron spectroscopy (HAXPES or HXPS) \cite{HAXPESvolume}, in which energies may go up to 5-20\,keV. 
However, no photoelectron diffraction measurements have as yet been carried out at 
these energies, although the extension of XPD to the hard x-ray regime is expected 
to open up additional analytical possibilities in accessing truly bulk properties of new materials \cite{fadley2005nima}. The aim of this paper is to assess some of 
these possibilities by means of accurate model calculations. 

In anticipation of the experimental realizations of such hard x-ray photoelectron diffraction (HXPD) experiments, it also seems clear that a modification of the theoretical approach will be necessary due to the much higher energies involved. The currently most often used method includes single and multiple scattering of photoelectrons within a finite cluster of atoms and allows the description of arbitrary surface structures which can possess short-range as well as long-range order. Such cluster calculations have been used successfully in a large number of experimental investigations \cite{kono80,osterwalder95,hayoz99,hardman99,schieffer01}, and in one case, even used to theoretically assess the emission from an adsorbate molecule for energies from 0.5\,keV to 10\,keV \cite{thompson1984jesrp}.
However, in dealing with multilayer substrate emission, it has been realized for some time that, for energies of about 1\,keV or more, the XPD patterns begin to show evidence of long-range Bragg-related effects that are known as Kikuchi bands \cite{goldberg80,trehan1987jesrp}. Such bands are well known, and in connection with a given low-index set of crystal planes (hkl) they are associated with enhanced intensity modulations over an angular width of twice the first-order Bragg angle of these planes:  $\sin\theta_{Bragg} = \lambda_e / 2d_{hkl}$, where $\lambda_e$ = the electron wavelength $\propto 1/\sqrt{E_{kin}}$ and $d_{hkl}$ = the interplanar spacing.  
Cluster simulations have in fact been successful in reproducing the formation of Kikuchi-bands in multilayer substrate XPD measured with high angular resolution \cite{kuettel1994ss,bardi97}. In these studies, it was shown that these bands become more pronounced as the number of scatterers in a cluster becomes larger.
Such cluster calculations have been made more time-efficient and accurate through different approximations for the spherical-wave scattering that is the natural starting point for photoelectron diffraction, since the initial wave is emitted from a localized core level \cite{chen98,saldin93,fritzsche86,abajo2001prb}.
However, because all of the cluster methods solve the scattering problem essentially in a \emph{spherical-wave} expansion of finite order, they become computationally more and more demanding at higher energies and for clusters with
a larger and larger numbers of atoms.
For example, for scattering of 10\,keV electrons from a representative atomic potential of \mbox{1\,\AA\,\,radius}, the number of scattering phase shifts has to be increased to $l_{max}\approx k r_{MT} = 100$, where $k$ is the electron wave vector and $r_{MT}$ is the muffin-tin radius of the atom.  
This poses considerable calculational difficulties as the photoelectron kinetic energies become larger than approximately 1\,keV. Beyond this, the clusters sizes needed at these energies also become larger due to the greater inelastic mean free paths, which roughly go as $E_{kin}^{0.75}$ and can be as large as 5-15\,nm for 10\,keV energy. The resulting clusters can thus easily
 contain several thousand atoms, and for example, about 10000 atoms in 40 atomic layers were needed in cluster calculations on Cu at about 1\,keV to be able to see the onset of Kikuchi band behavior \cite{trehan1987jesrp}.
This can make multiple scattering cluster calculations at high energies prohibitively time-consuming, as we have in fact verified as part of this study.

\begin{figure}\centering
  \includegraphics[width=14cm]{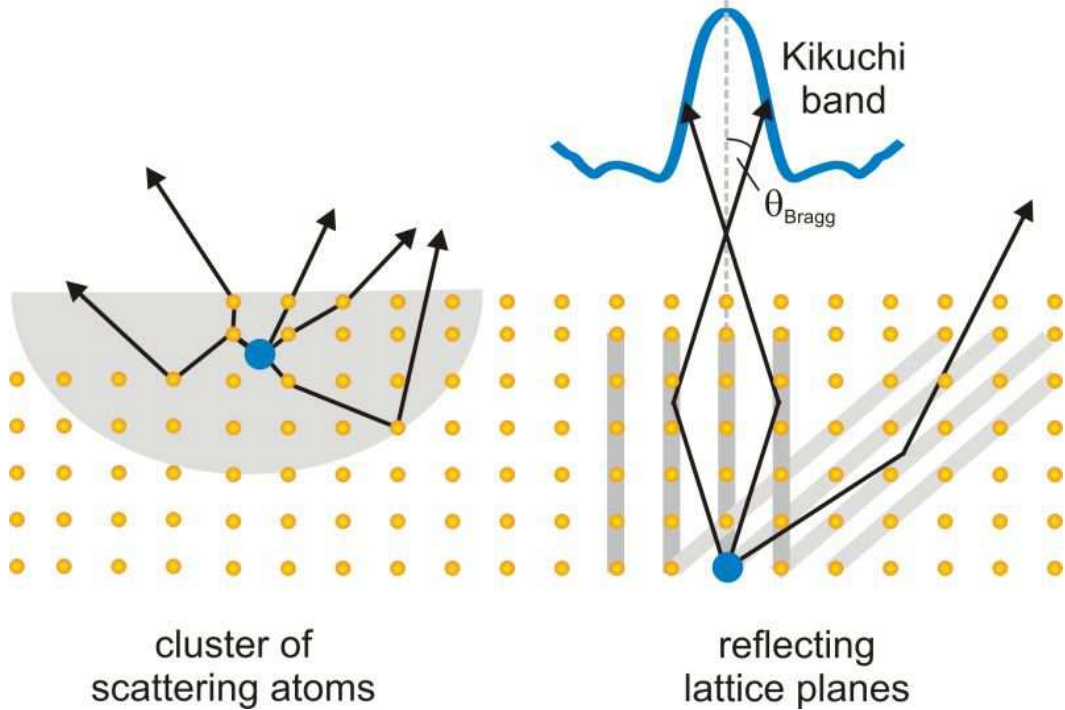}\\
  \caption{Electron diffraction from localized sources (blue):
  left: scattering atoms in a finite cluster around the source and no restrictions on symmetry, right: scattering by lattice planes of a long-range periodic structure }\label{clusterkiku}
\end{figure}

In order to frame the modelling in a manner more natural for multi-layer emission of such high-energy photoelectrons, we note that they sample not only the immediate surface layers where the bulk symmetry is broken, but also the diffraction patterns are formed from thicknesses on the order of the inelastic mean free paths which can be 5$\ldots$15 nanometers in the hard x-ray regime.
On these length scales, the properties of the uppermost surface layers play a lesser role relative to the increased contributions of the bulk crystal. 
In this case, photoelectron diffraction can be sufficiently well described by a theory which explicitly exploits the three-dimensional translational symmetry of the crystal scattering potential, with appropriate matching via suitable boundary conditions to the photoelectron wave function outside the surface.

In Fig. \ref{clusterkiku}, we show a schematic comparison of both of these  theoretical models of electron diffraction at surfaces:  the cluster picture (left) and the picture of reflecting lattice planes in a long-range periodic structure (right). While the cluster picture has the advantage of being able to in principle describe an arbitrary surface structure, as is indicated by the contraction of the first layer and a step edge, the Kikuchi bands are most easily explained by assuming the lattice planes of the periodic crystal as the fundamental scattering entities. For perfect crystals, both approaches are equivalent when taken to infinite order.

It is interesting to note that already the earliest observations of photoelectron diffraction on single crystal surfaces in the ca. 1keV regime have interpreted the angular distributions as being caused by reflection of the photoelectrons on lattice planes of the three-dimensionally periodic bulk crystal \cite{siegbahn70,fadley1971pla}. In this view, XPD is a special case of emission from point-source core levels inside a crystal, for which a general description can be obtained via a dynamical many-beam theory involving a plane wave expansion of the diffracted electron waves \cite{dewames1968aca}.
For example, a simple two-beam dynamical theory was applied to explain the azimuthal variations of photoelectron intensities for single crystal copper \cite{goldberg80}. Using this theory, the intensity variations could be reproduced by the summation of a number of  Kikuchi bands. These bands show increased intensity in a region of a width which is about twice the Bragg angle of the corresponding reflecting lattice plane, as qualitatively illustrated at right in Fig. 1.

Making use of this view of an effectively three-dimensional crystal that is diffracting the photoelectrons, one of the authors has previously been able to show that the Bloch wave approach of the dynamical theory of electron diffraction is able to provide a practical framework for the simulation of XPD patterns at high kinetic energies \cite{winkelmann2004prb}.
It was shown that this theory can not only explain the formation of Kikuchi bands but also reproduces features like forward-scattering directions and ring-shaped interference maxima around these directions.
 Such features can also be explained in the intuitive forward scattering picture of 
 cluster calculations, as has been shown by direct comparison to single scattering 
 cluster simulations \cite{winkelmann2004prb}. These rings are the long-range-order analogues of what have been termed first-order diffraction fringes in XPD studies of small molecules and chains \cite{fadley:synchrotron}.  
 The advantage of dynamical diffraction theory for substrate XPD lies in the fact that the lattice planes which cause the Kikuchi-lines are already an essential building block of the theory and do not emerge as a secondary phenomenon from a huge number of periodically arranged individual scatterers. In this way, the increasing number of atoms that is involved in the diffraction process does not necessarily imply a corresponding increase in the complexity of the interpretation of the results. 
Instead, as is well known, the diffraction process in a periodic medium is more conveniently calculated via a reciprocal space method.

The purpose of this article is to extend many-beam dynamical simulations to XPD
at hard x-ray energies, to assess what should be observed in experiment, and to point out some promising future applications of HXPD. We use primarily the examples of diamond C(111) and Si(111), for which 
there exist published experimental data at about 1\,keV, but also show some calculations for Fe. After showing that we can reproduce the is experimental data for C and Si with our approach, we then increase the kinetic energy up to 20\,keV and we analyze qualitative and quantitative features of photoelectron diffraction at these energies. 
We also assess the sensitivity of HXPD to two types of structural variation: a tetragonal lattice distortion and emission from an impurity or dopant atom sitting on a lattice or interstitial site.  
 For comparison, we also show some results obtained using a cluster model \cite{abajo2001prb}. 
The experimental and theoretical data presented here should thus provide an excellent roadmap of the characteristic features in the patterns to be expected in future hard x-ray photoelectron diffraction experiments.

\section{Theory}

Although the full details of our long-range-order dynamical approach have been presented elsewhere\cite{winkelmann2004prb}, we introduce the essentials briefly here.  
For the description of the photoelectron diffraction process, we assume in first approximation that the electrons originate isotropically from point sources which are periodically arranged inside a crystal.
Thus, we neglect any anisotropy associated with the photoelectric cross section. 
By making this approximation we obtain 
conclusions on the general features of the photoelectron diffraction process at
high energies, independent of a specific experimental geometry
and surface under consideration. 
Future improvements going beyond this approximation that would be necessary for a quantitative description 
of specific experimental results will be discussed below.

The outgoing photoelectrons are scattered by the crystal and detected as a plane wave at the detector. 
By using the reciprocity principle one realizes that the diffraction from a point source inside a crystal is equivalent to the problem of an incoming electron beam which is diffracted by the crystal and results in a certain electron intensity at the emitting atoms positions \cite{dewames1968aca}. 
The diffraction of an electron beam impinging on a sample is of obvious 
importance in various methods of scanning and transmission electron microscopy, which permits using the theoretical approaches that are applied in these methods. 
This close analogy to incoming beam diffraction effects allowed us to use existing Bloch wave algorithms developed for high energy electron diffraction \cite{humphreys1979rpp,emd} for this application to substrate XPD. 
We will only give a short summary of the general approach 
we apply. More details can be found in \citeauthor{winkelmann2004prb}\cite{winkelmann2004prb,winkelmann2007um}.

What we need to describe is the relative intensity distribution of photoelectrons
in the detected directions. The corresponding range of 
outgoing wave vectors is denoted by $\mathbf{k}^{\alpha}_{out}$ where $\alpha$ 
labels the outgoing direction in a suitably chosen 
coordinate system.
The high energy wave function inside the crystal is described as a superposition of Bloch waves 
with wavevectors $\mathbf{k}^{(j)}$:
\begin{equation}\label{wave}
\Psi(\mathbf{r})=\sum_j c_j \exp( i \mathbf{k}^{(j)}\cdot \mathbf{r}) \sum_{\mathbf{g}} C_{\mathbf{g}}^{(j)} \exp(i \mathbf{g}\cdot \mathbf{r}) 
\end{equation}
For a specific $\mathbf{k}^{\alpha}_{out}$ outside the crystal, we can determine  the electron wavevector $\mathbf{K}$ inside the crystal which is used to express $\mathbf{k}^{(j)}$ as
$ \mathbf{k}^{(j)}=\mathbf{K}+\lambda^{(j)}\mathbf{n}
$ where $\mathbf{n}$ is a unit vector normal to the surface. Starting from the Schr\"odinger equation, one can then set up an eigenvalue problem \cite{emd} which gives the eigenvalues $\lambda^{(j)}$ and eigenvectors with elements $C_{\mathbf{g}}^{(j)}$. 
The scattering potential is described in terms of its Fourier coefficients, with a constant real part to represent the surface or inner potential $V_{0r}$ and a constant imaginary part to represent the electron inelastic mean free path $\Lambda_e$ through $eV_{0i}=-\sqrt{\hbar^2 E_{kin}/2m_e}/\Lambda_e$.

The boundary conditions at the surface determine the coefficients $c_j$ in (\ref{wave}). 
After this, the wave function (\ref{wave}) is known and can be used to calculate the probability density inside the crystal for a plane wave moving in the $\mathbf{k}^{\alpha}_{out}$ direction. 
The modulation of this probability at the photoemitter positions describes the diffraction
by the crystal. To couple the photoemitters to the Bloch wave field, we need
the corresponding matrix elements of the interaction operator $V_I$ which describes the photoexcitation due to the vector potential $\mathbf{A(r)}$ of the incident radiation \cite{schattkesspe}, $V_I \sim \mathbf{A(r)\cdot p}$,  with the momentum operator $\mathbf{p}$ .
The matrix element needs to be determined between an atomic core level state $\ket{L}$ characterized by quantum numbers L=(l,m,...) 
and the plane waves $\ket{\mathbf{K,g}}$ of the wave function (\ref{wave}).

In a general way, we can write the dynamical diffraction part of the cross-section of localized scattering processes as \cite{rossouw94}:
\begin{equation}
I_{DYN}\propto \sum_{i,j} B^{ij}(t) \sum_{\mathbf{g},\mathbf{h}} C^{(i)}_{\mathbf{g}}~C^{(j)*}_{\mathbf{h}}~ \mu_{\mathbf{g},\mathbf{h}} 
\label{dyn}
\end{equation}
with a depth integrated interference term $B^{ij}(t)$ of the Bloch waves $i$ and $j$:
\begin{equation}
    B^{ij}(t)=c_i~ c_j^* ~\frac{\exp[ i(\lambda^i-\lambda^{j*})t]-1}{ i(\lambda^i-\lambda^{j*})t}
\end{equation}
Here, the terms $\mu_{\mathbf{g},\mathbf{h}}$ contain products of the matrix elements of $V_I$ using summation indices $\mathbf{g}$ and $\mathbf{h}$ for the plane wave expansion. 
For this work, we will assume that photoemission takes places isotropically from the
atomic positions  $\mathbf{r}_n$ of the emitters, which are described via delta functions $\delta(\mathbf{r}-\mathbf{r}_n)$ broadened by thermal vibration according to a Debye-Waller factor. This leads to coefficients of the form 
$\mu_{\mathbf{g},\mathbf{h}}=  \exp(-M_{\mathbf{g}-\mathbf{h}}^n) \exp[ i (\mathbf{g}-\mathbf{h})\cdot \mathbf{r}_n ]$
and  results in the cross section formula which we have used in our previous simulations for quasi-elastic backscattering of electrons and XPD \cite{winkelmann2004prb,winkelmann2007um}:
\begin{equation} \label{rossouwbse}
    I_{DYN} \propto \sum_{n,ij} B^{ij}(t) \sum_{\mathbf{g},\mathbf{h}} C_{\mathbf{g}}^{(i)} C_{\mathbf{h}}^{(j)*} \exp(-M_{\mathbf{g}-\mathbf{h}}^n) \exp[ i (\mathbf{g}-\mathbf{h})\cdot \mathbf{r}_n ]  
\end{equation}

Our assumption of simple isotropic photoelectron emission as compared to a more correct description of the ionization process is made to gain insight into the fundamental processes of dynamical diffraction by lattice planes in photoelectron diffraction. This assumption also leads to results that are directly applicable to the relative changes in intensity seen for emission from a given orbital for the often-occurring case in which the photon-electron geometry, and thus the differential photoelectric cross section, are fixed and the sample is rotated in polar and azimuthal angles so as to scan the emission direction in principle over the full hemisphere of emission, as is shown e.g. in Fig. \ref{expsim}.  
For a future more quantitative description of specific experiments, including intensity ratios between different core levels, this theoretical description can be extended to include the details of the photoemission process. 
This concerns especially the matrix-element effects of the $l\pm1$ channels at the ionization. These can be incorporated into the theory by the use of generalized potentials $\mu_{\mathbf{g},\mathbf{h}}$ which contain the necessary dipole matrix elements for inner shell atomic excitation \cite{allen90,allen93,saldin1987pmb}.
In principle, this simply amounts to a plane wave expansion of the photoexcited states described by the $l\pm1$ quantum numbers in the atomic spherical wave description \cite{daimon1995jesrp}.

Another approximation used in our approach concerns the neglect of backscattering, which should be valid at high energies for which electron-atom scattering is known to be strongly peaked in the forward direction. Inclusion of backscattering in the theory results in eigenvalue problems of size 2$N$x2$N$ as compared to $N$x$N$ when neglecting backscattering, where $N$ is the number of 
included Fourier coefficients of the crystal potential \cite{qian1993aca}. We expect that backscattering will have a higher effect at lower energies and along directions with strong multiple scattering, e.g. zone axis directions.

\section{Results}
In this section, we will first show that the dynamical electron diffraction
approach is able to reproduce published experimental data for XPD at kinetic energies near 1\,keV from diamond C(111) and from silicon Si(111) with very good agreement.
We then calculate the corresponding XPD patterns at kinetic energies ranging from 0.5 to 20\,keV for comparison.

\subsection{XPD from Diamond and Silicon near 1\,keV}

\begin{figure}\centering
  \includegraphics[width=14cm]{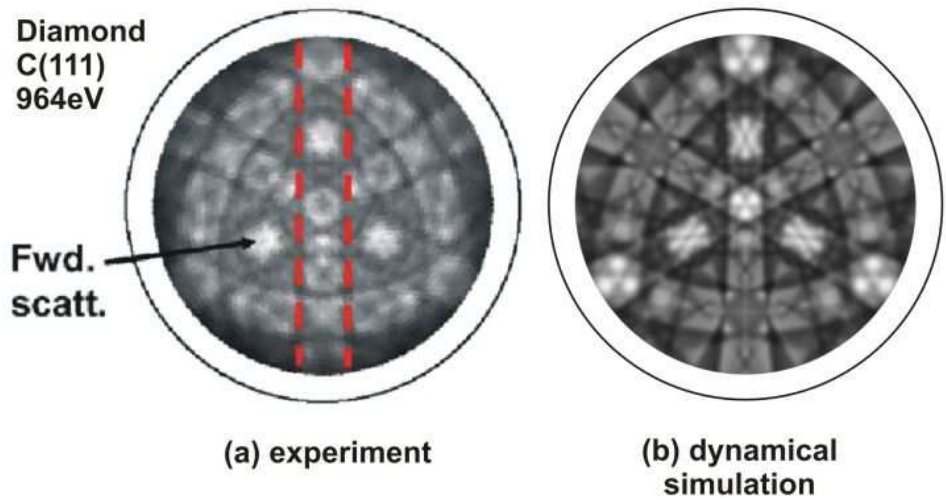}\\
  \caption{(a) Experimental data from \citeauthor{osterwalder1994jesrp} \cite{osterwalder1994jesrp,kuettel1994ss}, (b) simulated photoelectron diffraction from diamond C(111) at a kinetic energy of 964\,eV. In the simulation, 97 Fourier coefficients ("reflecting lattice planes") have been taken into account. }\label{expsim}
\end{figure}

\begin{figure}\centering
  \includegraphics[width=14cm]{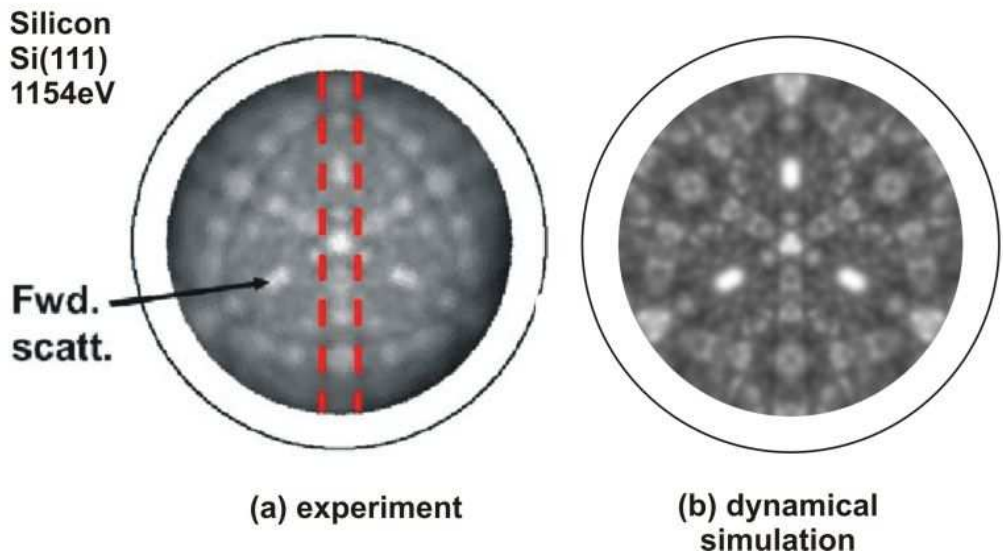}\\
  \caption{(a) Experimental data from \citeauthor{osterwalder1994jesrp} \cite{osterwalder1994jesrp,kuettel1994ss}, (b) simulated photoelectron diffraction  photoelectron diffraction from
  Si(111) at a kinetic energy of 1154\,eV. In the simulation, 127 Fourier coefficients have been taken into account. }\label{expsimsi}
\end{figure}

In Fig. \ref{expsim}a, we show the experimental data of \citeauthor{osterwalder1994jesrp} \cite{osterwalder1994jesrp,kuettel1994ss} for C1s photoemission in diamond C(111) observed at a kinetic energy of 964\,eV.
In part b of Fig. \ref{expsim}, the result of an XPD simulation using dynamical electron diffraction theory for point sources in a crystal is shown.
We assumed a lattice constant of 3.57\AA~ and isotropic emission of photoelectrons from the C atoms. The 97 strongest Fourier components of the lattice potential have been taken into account \cite{winkelmann2004prb}. These correspond to reciprocal lattice vectors ($h$,$k$,$l$) associated with various reflecting lattice planes, including possible higher order reflections. The calculation has been averaged over a sufficient angular range to account for the limited experimental resolution.

There is excellent agreement between the experimental data and the simulation in Fig. \ref{expsim}. In particular, the Kikuchi-band structure is correctly reproduced and permits interpreting weak features in the experimental data like the central decrease of intensity in the marked forward scattering directions in Fig. \ref{expsim}a.
This is obviously due to a crossing of dark Kikuchi lines and leads to the well
known volcano shape in XPD polar plots. As we have already noted, it is likely that probably much more computational effort would be needed in a cluster approach to reach this level of agreement, but we return later to show some calculations using this approach for completeness. While a simple geometrical analysis of Kikuchi line positions will verify the fact that there are indeed lines crossing in this region at this energy, this observation does \emph{not} give reasons why they should be dark. We really need to consider the dynamical diffraction process to account correctly for the observed \emph{intensities}.

A similar degree of agreement can be seen for emission from silicon Si(111) at 1154\,eV shown in
Fig. \ref{expsimsi}, although the larger lattice constant of  5.43\AA~ means the features are inherently narrower, and it is a challenge for the experimental data to show all of the fine structure present in theory . For the theoretical pattern, 127 Fourier components
have been included. In these results, both for experiment and theory, we recognize the appearance of ring structures around low-index emission directions which can be interpreted as first order interference maxima around the corresponding central forward-scattering direction \cite{fadley:synchrotron}. It has been shown before that these features can be translated into the language of high energy electron diffraction, where they are named "higher order Laue zone" (HOLZ) rings \cite{winkelmann2004prb}. In both pictures, these features consist of intensity which is scattered away from the forward scattering direction. The diameter of the ring is determined by the separation of scatterers along the forward-scattering direction. This fact can be exploited to obtain 
a holographic type of information from these rings 
\cite{osterwalder1994jesrp,michael2000um,fadley2001jpcm}.
This line of thought leads to the conclusion, that if these rings are created by electrons that are scattered \emph{away} from their initial direction in the forward scattering peak, then there has to be a reduction of intensity near this particular initial direction. In the picture of scattering lattice planes: a Kikuchi line corresponding to a reciprocal lattice vector $\mathbf{g}$ is crossing the forward scattering peak, and there must be another Kikuchi line that corresponds to $-\mathbf{g}$ forming part of the envelope of the 
outward ring. By the diffraction process, intensity is transferred from the initial forward 
scattering direction (related to $\mathbf{g}$) to the outer ring
formed as the envelope of similar $-\mathbf{g}$ reflections. While the outer ring 
lines are high in intensity, the initial forward scattering lines are dark. This can be actually seen in the center of the rings marked in the simulated Si pattern of Fig. 
\ref{expsimsi}. We will demonstrate below
that a rather complicated fine structure of dark lines exists at high kinetic 
energies and that this is very sensitive to the crystal geometry.
From this fine structure, valuable information should be obtainable in a way which is qualitatively different from the low energy regime.

Summarizing this section, within our simple assumption of isotropic emission of photoelectrons, the one-to-one agreement of features in the simulation and experiment is most encouraging. This permits making semi-quantitative estimates concerning the effects that will be seen at even higher energies.

\subsection{Simulations of high energy XPD}

\subsubsection{Energy dependent full-hemispherical patterns}

\begin{figure}
\centering
  \includegraphics[width=14cm]{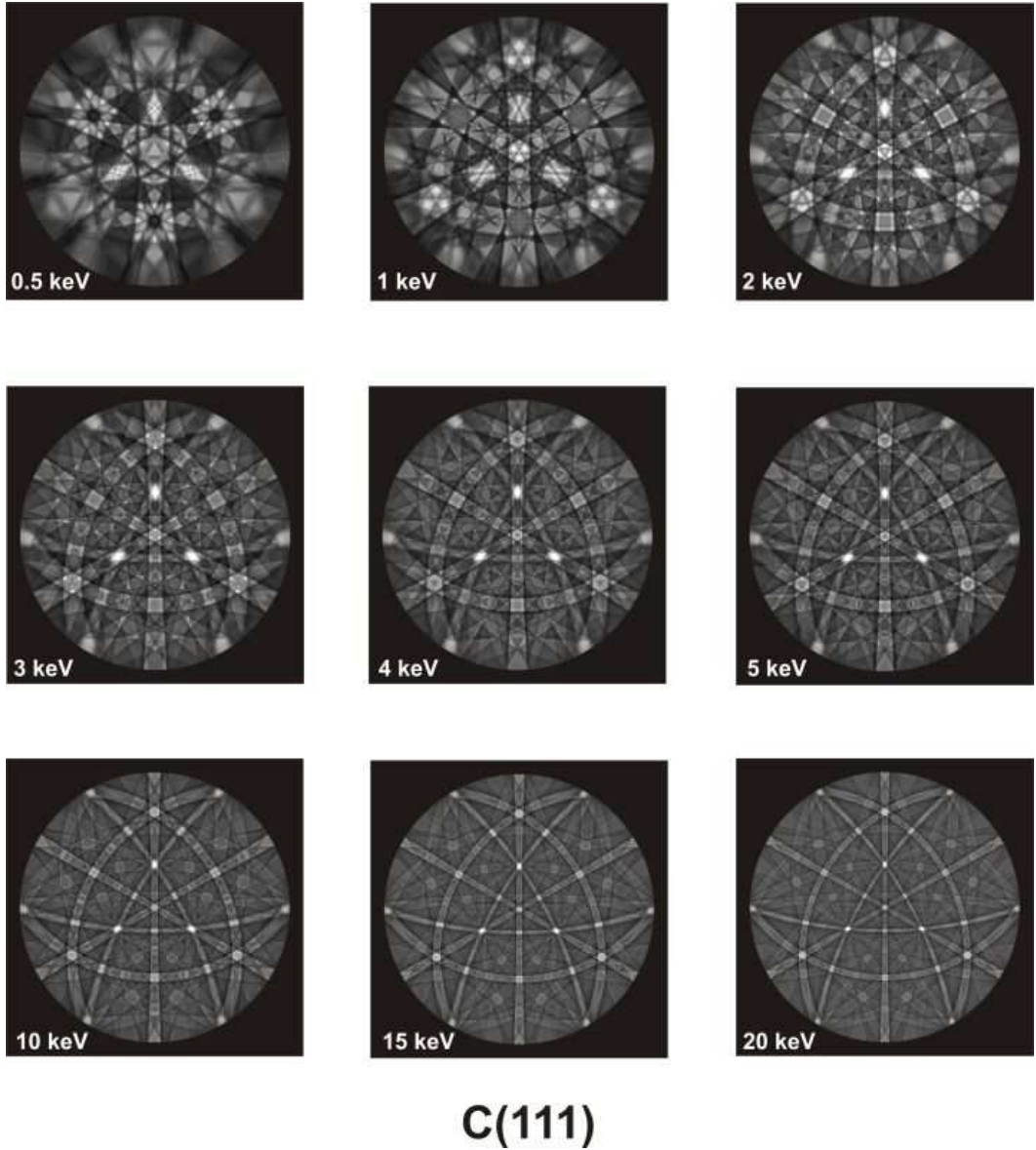}
  \caption{Simulated photoelectron diffraction patterns from diamond C(111) at the kinetic energies indicated in keV. The circular areas correspond to the whole hemisphere above the sample in stereographic projection. The same parameters as in Fig. \ref{expsim} 
have been used.
}
\label{sim0520C}
\end{figure}

\begin{figure}
\centering
  \includegraphics[width=14cm]{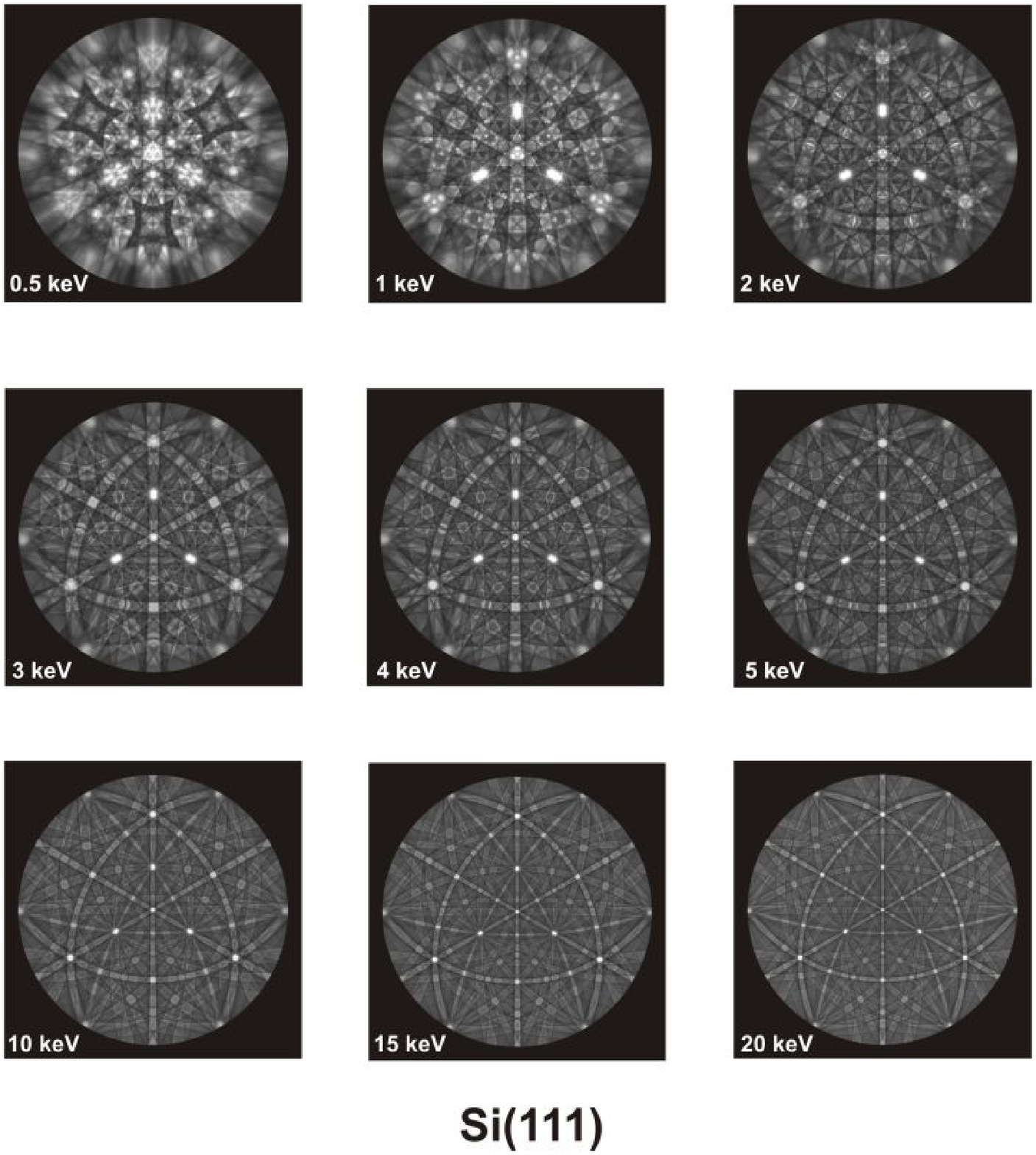}
  \caption{Simulated photoelectron diffraction patterns from silicon Si(111) at the kinetic energies indicated in keV. The circular areas correspond to the whole hemisphere above the sample in stereographic projection. The same parameters as in Fig. \ref{expsimsi} have been used, with the mean free path adjusted with energy.
}
\label{sim0520Si}
\end{figure}

Having convinced ourselves that our calculational approach is able to describe 
the XPD patterns of diamond and silicon at around 1\,keV kinetic energy, we 
will now show the results of calculations at systematically higher energies up to 20\,keV. In Figures \ref{sim0520C} and \ref{sim0520Si} we show calculations for the same two cases C(111) and Si(111), respectively, 
starting from 500\,eV and assuming perfect angular resolution of the electron analyzer.
The energy values are spaced in such a way as to see on which energy scales there are significant changes.

At the lowest energies, we see a relatively rapid variation of the XPD patterns with energy  e.g. from 0.5\,keV to 1\,keV to 2\,keV. The appearance of very clear Kikuchi lines at 1\,keV can be noticed, while at lower energies the larger Bragg angles lead to  features of correspondingly larger angular extension for which the Kikuchi band character of the patterns is more difficult to discern. Still, a network of relatively sharp and dark Kikuchi lines remains.

With increasing energy, the geometrical pattern of narrowing Kikuchi bands entirely dominates the character of the pattern. This is because the angular range on which intensity changes can happen due to diffraction is roughly determined by the size of the Bragg angles. These decrease with with increasing energy, so that the intensity variations due to the multiple (dynamical) scattering occur in a more and more limited region around the projected lattice planes in the patterns. Correspondingly, the crystallographic symmetry of the crystal lattice planes becomes the defining factor of the pattern structure. At the same time, the patterns
vary less and less as a function of energy due to the the energy dependence of the Bragg angle $\theta$: $\sin \theta \propto 1/\sqrt{E}$ for a given reflection $\mathbf{g}$.  

While we can be very confident that the high energy patterns starting from 1\,keV give a realistic estimation of the XPD data to be expected, we have to stress that the 0.5\,keV patterns can be taken only for qualitative comparison. Due to the approximations of our theoretical approach, it is expected to be increasingly unreliable for energies which are considerably below 1\,keV in the case of weaker-scattering low-Z materials like C and Si. At these energies, a cluster approach is probably more suitable.

\subsubsection{Simple interpretation of the photoelectron diffraction process at high energies}
\label{sec:blochdenXPD}

\begin{figure}
\centering
  \includegraphics[width=14cm]{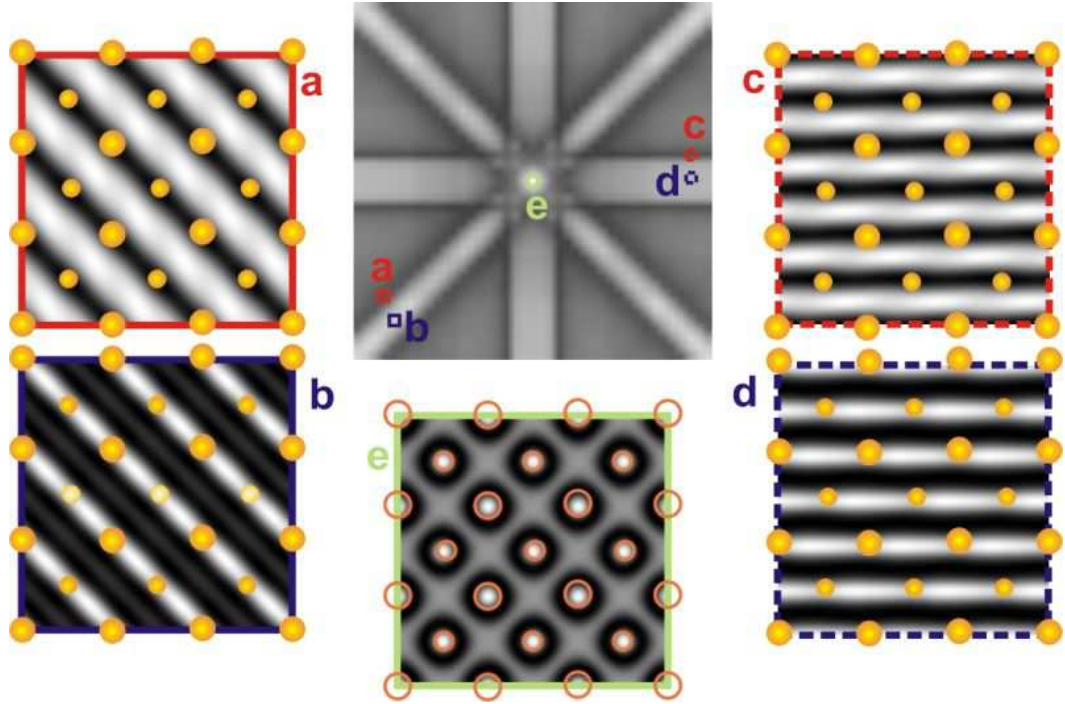}
  \caption{Top center panel: Simplified 9-beam Kikuchi pattern for an Fe bcc lattice at 15\,keV kinetic energy
     for an angular range of $\pm $ 15 degreees from the [001] surface normal.
     The areas marked by colored squares and letters a-e in the Kikuchi pattern
      each correspond to a specific detection direction of either high or low intensity. The other panels then show the  (x,y) probability density distribution in the surface plane corresponding to (a)-(e) for 3x3 unit cells averaged along the z-axis [001] (white=high probability of ending up in the plane wave to the  detection direction, black=low probability).
 The photoelectrons are created via localized excitations at the atomic positions (colored circles) and are diffracted into the detection directions with an intensity proportional to the overlap with the probability density of the diffraction process.
The central lower pattern (e, green border) corresponds to the forward scattering direction of the surface normal and shows how the symmetry of the lattice can confine the electrons to channel along the zone-axis direction. The results in the other panels (a)-(d) explain clearly the high or low nature of intensity along the associated directions.}
\label{blochdenXPD}
\end{figure}

We see from the simulation that the photoelectron diffraction patterns become
dominantly characterized by a network of Kikuchi bands as energy is increased. These have the
general property of being symmetric with respect to a projected lattice
plane and they appear either with increased intensity in the middle of the band
bordered by two lines of decreased intensity near the Bragg angle or they show decreased intensity in the middle bordered by higher intensity. In the cluster picture, it is not straightforward to explain this behavior. For a periodic crystal, by contrast, the mechanism  that leads to the intensity distribution in a Kikuchi line can be visualized and interpreted in a very direct way.

In the photoelectron diffraction Kikuchi pattern, we have directions of higher and of lower intensity. This results from the fact that the photoelectrons, which are created at localized positions within the unit cell, have a different probability to couple to the
plane wave which moves in vacuum into the detection direction. In the diffraction pattern, we sense the modulation of this coupling with changing detection direction.
We will see in the following that for changing detection directions,
different locations in the unit cell will couple with different efficiency to the outgoing plane wave. Because the sources of photoelectrons are at fixed positions, this leads to the effect, that only at certain detection angles is the diffraction process able to couple the photoemitted electrons with high probability into the outgoing wave. At the positions of dark Kikuchi lines, the photoelectrons would have to originate from positions
\emph{between} the emitter atoms, a condition which is obviously impossible to fulfill.  Alternatively, one can think about this process by using the reciprocity principle and looking at the probability density in the unit cell for plane waves coming from the detection directions. If this probability density is high at the photoemitters, a high photoelectron intensity is observed, and if it is high between the photoemitters, low intensity is observed.

These qualitative considerations are made visible by simulations in Fig. \ref{blochdenXPD} for 15\,keV emission from an Fe bcc lattice. In the central upper panel of Fig. \ref{blochdenXPD}, we show a calculated model Kikuchi pattern for electrons
emitted from bcc Fe(001) at 15\,keV. This is a simplified Kikuchi pattern, taking into account only 8 reflecting lattice planes of the \{200\} and \{110\} family which are oriented perpendicular to the surface plane. These lattice planes determine the 4 crossing Kikuchi bands in the pattern, each corresponding to a pair of reciprocal lattice vectors $\mathbf{g}$
and $\mathbf{-g}$ . In the Kikuchi pattern we have marked some specific detection directions corresponding to the middle and the border of Kikuchi bands. In the other panels of Fig. \ref{blochdenXPD}, we then show how the probability density for
an incoming plane wave from a specific detection direction is distributed inside the unit cells. We view the crystal along the direction of the [001] surface normal, and since the diffracting lattice planes are oriented perpendicular to the surface, we average the probability density
along the $z$-axis. The positions of the atoms are marked by circles, the small atoms are centered in the bcc unit cell.

In the unit cell panels of Fig. \ref{blochdenXPD}, we can immediately see how 
the crystal symmetry is reflected in the probability intensity distributions in accordance with the orientation of the reflecting lattice planes.
In the left upper and lower panels, we see that the photoelectron intensity
is high at the Kikuchi band position marked by the blue solid square (b) because the probability density of diffraction into the detection direction is also high at the
positions of the emitting atoms.
At the position of the dark Kikuchi line, marked by the red solid square (a), we see
that the probability density is concentrated \emph{between} the photoemitters, so that
it is more unlikely for photoelectrons to be diffracted in the direction of the red square, which is then manifested by the lower photoelectron intensity. The same reasoning holds for the other lattice planes, e.g. the directions marked by the dashed squares (c,d) shown on the right of Fig. \ref{blochdenXPD}, where we see that the variations in the probability density conform to the lattice symmetry by fitting into the space between or along the atomic positions (e.g. parallel to the (100) planes in the right part, or parallel to the (110) planes in the left part of Fig. \ref{blochdenXPD}).

More complicated intensity distributions arise near the directions where two or more lattice planes cross. As an example we show the forward scattering direction
along the [001] surface normal in the panel with the yellow border. We see that the
perfectly symmetric arrangement of the (100) and (110) lattice planes considered  
confines the probability density along the atom columns, which corresponds to 
electrons channelling along the atomic rows.

In the general case, the scattering lattice planes can be arbitrarily oriented (e.g. tilted with respect to the surface plane), and one would have to look at the
three-dimensional probability density within the unit cell, without averaging along $z$. 
While this certainly would give a more realistic picture, the basic mechanism is completely conveyed already by the simplified pattern of Fig. \ref{blochdenXPD}. 

In summary, we have supplied a visual interpretation of the relative variation of  the photoelectron diffraction intensities in Kikuchi features by 
plotting how the probability density of an \emph{incoming} plane wave is diffracted to different parts of the crystal unit cell and how it overlaps with the
photoemitters to varying degrees.

\subsubsection{Experimental implications and possible applications}

Based on the simulations above, we consider some implications for future HXPD
experiments. First of all, this concerns the angular width of the features to be observed and the necessary angular resolution.

\paragraph{Angular resolution}

\begin{figure}
\centering
  \includegraphics[width=14cm]{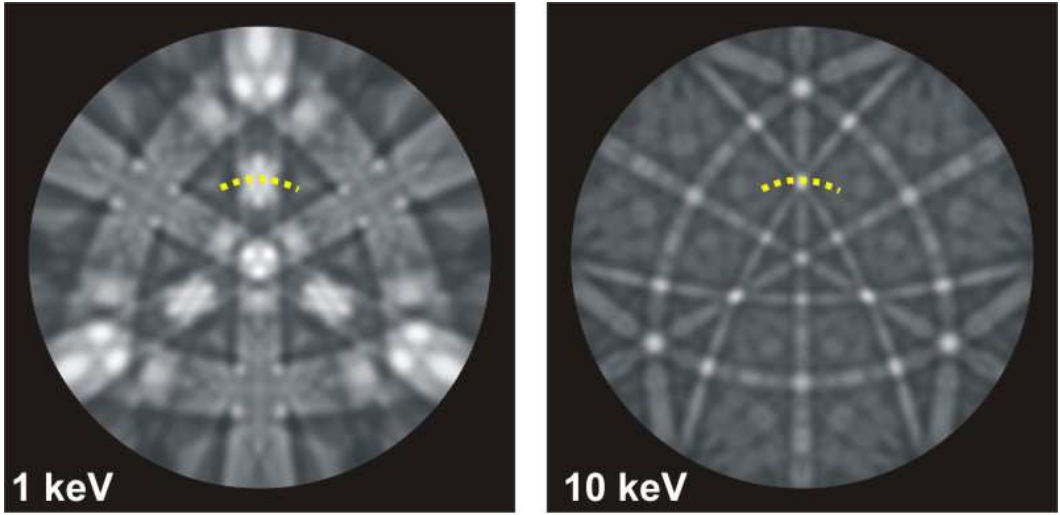}
  \caption{Direct comparison of simulated photoelectron diffraction patterns from diamond C(111) at 1\,keV (left) and at 10\,keV (right) with an angular resolution similar to the experimental pattern in Fig. \ref{expsim}.}
\label{reso}
\end{figure}

\begin{figure}

\centering
  \includegraphics[width=14cm]{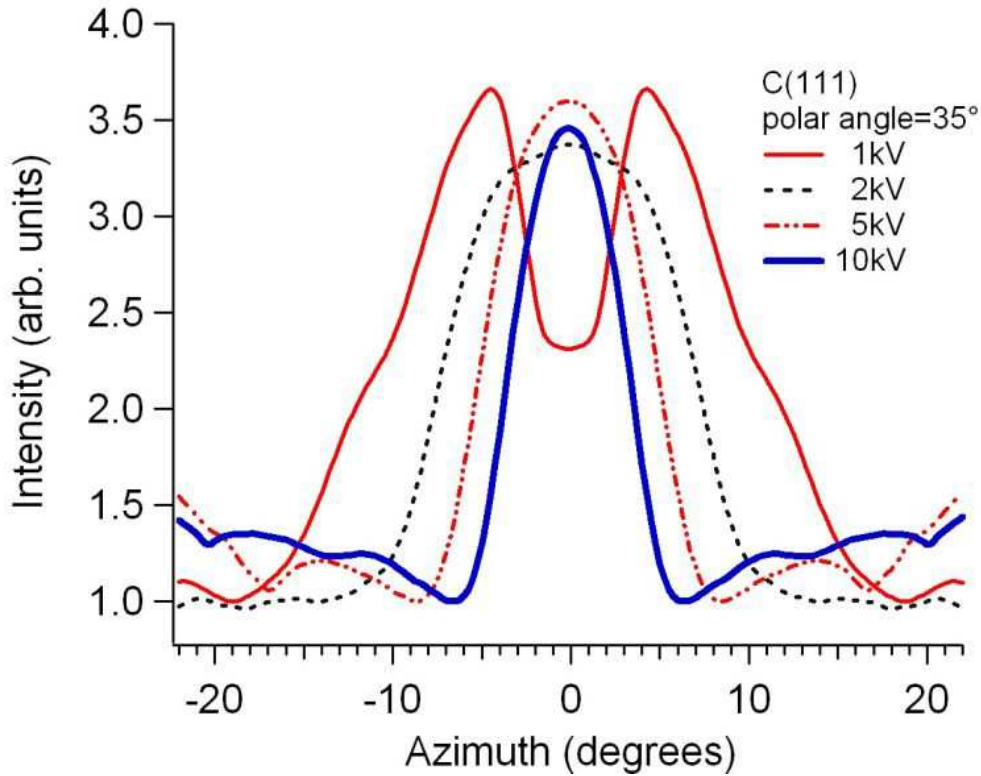}
  \caption{Calculated azimuthal plots for C(111) at various energies near the forward scattering peak at a polar angle of 35 degrees. The angular range corresponds to the dotted lines shown in Fig. \ref{reso}. An angular resolution of 
$\pm$1 degree has been assumed.
}
\label{aziplot}
\end{figure}

In figure \ref{reso}, we compare 
two simulated HXPD patterns of C(111) which are blurred according to an angular 
resolution of $\pm$1 degree (comparable to resolutions used in existing XPD 
experiments). One immediately notices the strongly reduced angular range over which the 
variations occur at 10\,keV photoelectron energy. To quantify this further, we show in Figure \ref{aziplot} an azimuthal scan over the regions indicated by dotted lines in Figure \ref{reso} at energies of 1, 2, 5, and 10\,keV. The 10\,keV peak is only about one third of the width of the 1\,keV peak, which is in agreement with an estimation from Bragg's law which would give a value of $1/\sqrt{10}=0.32$ for the change in $\sin \theta$.  We also notice that the peak intensity relative to the Kikuchi band minimum  
is reduced by less than 10\% when going from 1\,keV to 10\,keV, which indicates that such effects should be readily measurable. However, a spectrometer using  a larger angular averaging (lower angular resolution) would necessarily selectively reduce the peak-to-background ratios in the HXPD features due to their smaller angular widths.  We also note that Figs. 4 and 5 suggest a possible tradeoff between the angular resolution necessary to make make use of such structure and the energy.  That is, it may be useful to work at somewhat lower energies in the 2-5\,keV range in order to better match the size of the Kikuchi band structure to the angular resolution of the spectrometer.

As we have already remarked above, the 1\,keV peak in the azimuthal plot shows a well known volcano shape, whose origin can be interpreted by our simulation in a rather simple way as being due to a crossing of dark Kikuchi lines (see the discussion of the simulated C(111) data in \mbox{Figure \ref{expsim}}).

\begin{figure}
\centering
  \includegraphics[width=14cm]{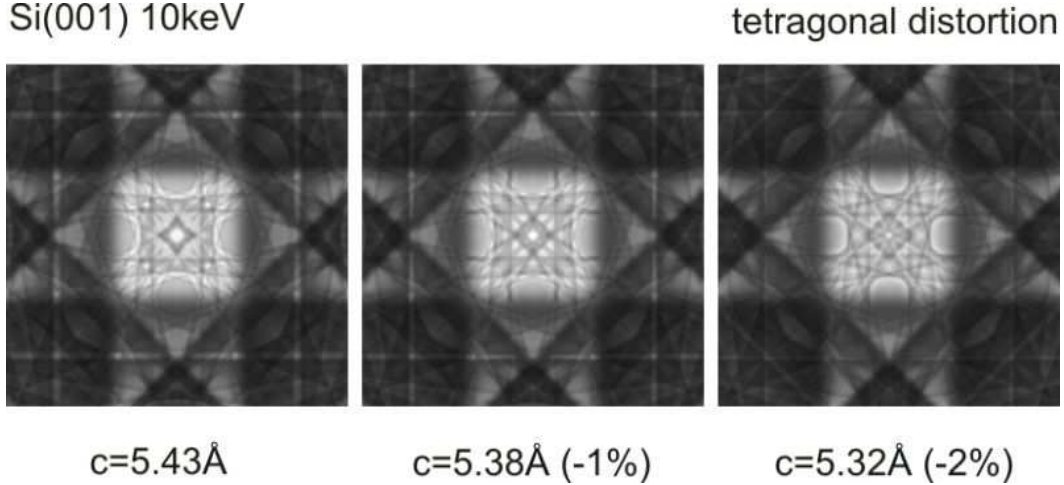}
  \caption{Calculated photoelectron intensity $\pm$5 degrees away from the surface normal of
  Si(001) at 10\,keV showing fine structure which is very sensitive to a possible   
tetragonal distortion. The left panel corresponds to the cubic Si unit cell, for the other panels, the unit cell has been compressed parallel to the surface normal along the c-axis by 1\% (middle) and by 2\% (right). Changes are clearly noticeable at 1\% distortion.}
\label{Si001tetra}
\end{figure}

\paragraph{Sensitivity to lattice distortions: tetragonal example} 
We have demonstrated above that the high energy XPD patterns are dominated
by the crystallographic symmetries. The angular scales on which  relevant changes
are visible are considerably reduced. This also means that it might not be necessaryily useful in all cases to measure large angular ranges at these high energies. A different approach at high energies would be to measure relatively confined angular ranges of the diffraction pattern with highest angular resolution, as might be provided in the future by the further development of special types of display analyzers \cite{kroemker2008rsi,daimon2006pss}. In fact, contemporary hemispherical electrostatic analyzers that are used for valence-band angle-resolved photoemission (ARPES) can already achieve angular resolutions in the 0.1 degree range over an angular interval of $\pm$ 10-20 degrees that would span the features shown in Fig. \ref{Si001tetra}. 
In the following we will show that useful structural information should be obtainable by measuring the details of the diffraction pattern in a region near a forward scattering direction (zone axis).

An important problem for which this approach might be useful is the growth of epitaxial thin films on a flat single-crystal and their possible tetragonal distortion.
In XPD, tetragonal distortions can be sensed by the change in the direction of forward scattering peaks. This approach of course remains valid at high energies.
However, to show that HXPD could give important results on this type of question 
also in a different mode of measurement, we calculated theoretical photoelectron 
diffraction 
intensities with high angular resolution near the [001] zone axis direction of 
Si(001). The idea is to
use information in the complex fine structure of a peak in a reduced angular area, rather than to compare 
two peak positions which are relatively far apart.

The patterns shown in Figure \ref{Si001tetra} correspond to an angular region of about $\pm$5 degrees around the Si(001) surface normal. To get an impression
about the relative size of this area, it is comparable to the central white spot in the calculated Si(111) pattern for 10\,keV in \mbox{Fig. \ref{sim0520Si}}.
We calculated the patterns
for electrons at 10\,keV kinetic energy in cubic Si and for films compressed by 1\% and 2\% along the direction of the surface normal. A tetragonal distortion of 1\% can be easily sensed, with a distortion of 2\% being even more obvious. By means of calculations in a small energy range around 10\,keV, we also estimate that a even relatively modest energy resolution of approximately 50\,eV would be sufficient to observe the effect, thus making such measurements much faster. This simplification in the measurement is because the 50\,eV change in energy changes the electron de Broglie wavelength by only $\sqrt{10050/10000}=1.0002$.  
In this sense, one could trade lower energy resolution for higher angular sensitivity in a display analyzer.

\paragraph{Impurity or Dopant Site Determination}

Another possible application of high energy XPD concerns the determination of
the crystallographic position of impurities in an otherwise perfect crystal.
A similar problem appears in the context of fluorescence radiation from impurities under excitation of crystals by either x-rays \cite{batterman1969prl} or electrons \cite{tafto1982um} while changing the incidence angle.
Close similarities also exist to the emission channelling technique using
high-energy conversion electrons from implanted radioactive isotopes \cite{wahl1997prl}.

\begin{figure}[b]
\centering
  \includegraphics[width=14cm]{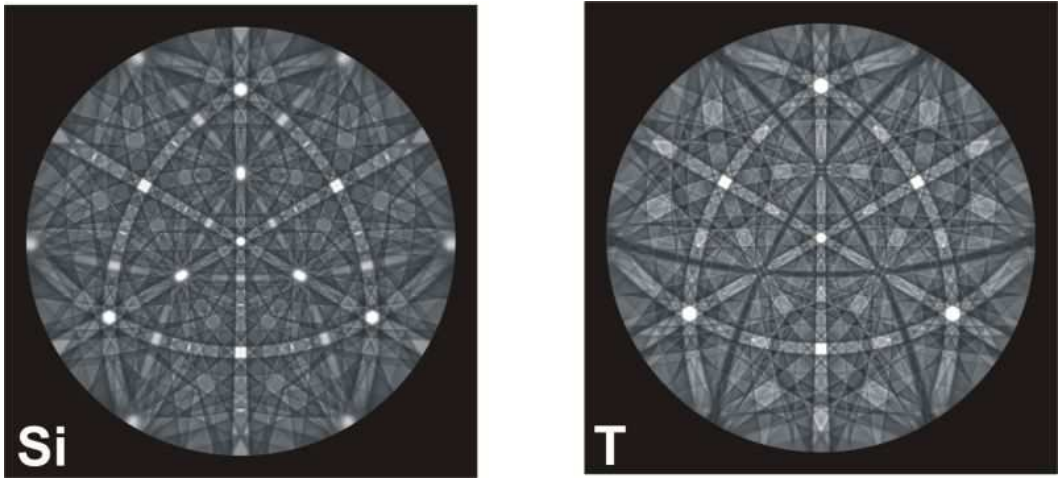}
  \caption{Calculated photoelectron diffraction pattern for a Si(111) surface at 6\,keV photoelectron kinetic energy from impurities statistically distributed  in the Si lattice. left: impurity located at the substitutional (Si) sites, right: impurity located at the tetrahedral interstitial (T) sites.}
\label{impur}
\end{figure}    

By reciprocity we see that the process of sensing a diffracted incident electron intensity by \emph{atomic detectors} that are placed somewhere inside the unit cell (emitting 
fluorescence x-rays according to how many electrons hit the atom as a function of incidence angle) is in principle the time-reversed version of the excitation of
localized \emph{atomic sources} by x-rays, producing photoelectrons and detecting the diffraction along the outgoing path to the detector. As a consequence, detection of impurity sites
in HXPD should be possible via the same interpretational approaches as used in the electron microscopy technique of 
"atom location by channelling enhanced microanalysis" (ALCHEMI \cite{tafto1982um}). The basic mechanism of impurity site determination is again closely based on the diffraction of high energy electrons to specific sections of the unit cell as discussed above in connection with Fig. \ref{blochdenXPD}. Because of the element-specific photoelectron excitation, an impurity which is statistically distributed in the crystal and which takes a specific site in a unit cell can act as an identifiable detector of the diffracted intensity in that part of the unit
cell which it occupies, without at the same time taking part in the three-dimensional lattice periodicity which dictates the character of the diffraction process. This means that the impurity could also be placed
at an interstitial site and thus would show a Kikuchi line profile which is inverted with respect to the other atoms.
This leads to the general idea of identifying an impurity site by its characteristic Kikuchi band profile in the HXPD pattern.
In fact, it has been previously shown that site selective
Kikuchi patterns can be reproduced by our approach for the case of photoelectron diffraction from AlN and CaF$_2$ at about 1\,keV energy\cite{winkelmann2004prb}. 

This site specific behavior is expected to remain valid at higher energies, and we
illustrate these effects in Fig. \ref{impur}, where the HXPD patterns of an impurity source in Si, emitting 6\,keV photoelectrons from either the substitutional (Si) site (left) or from the interstitial site (T) of tetrahedral symmetry (right) are compared.
Fig. \ref{impur} clearly shows that characteristic dark bands appear in the HXPD pattern from the T-site as compared to the Si-site. This demonstrates that it should be possible to determine impurity or dopant sites in bulk crystals and buried films under hard x-ray excitation by their characteristic diffraction patterns in a long-range ordered host crystal. 

\paragraph{Effects beyond the inelastic mean free path picture}

Compared to the elastic interference effects that lead to changing Kikuchi band profiles of impurities occupying different positions in the unit cell, the qualitatively different contrast reversal effect of Kikuchi bands described below could allow measurements of localized absorption effects.
In photoelectron spectroscopy and diffraction, the usual way to account for the losses of photoelectrons from the elastic channel is to introduce an inelastic mean free path corresponding to a constant imaginary part of the inner potential $V_{0i}$.
This will reduce the electron intensity with the travelled distance independently of the electron's place and direction in the crystal. It can be easily imagined, however, that the probability of inelastic scattering can depend on whether the electron wave field has a maximum amplitude between or at the atomic positions. In the latter case, any inelastic process that is localized at the atomic positions should be increased. This behavior is not captured by a constant imaginary potential, which can only reproduce the isotropic absorption. Additional Fourier components of the imaginary potential can be introduced to describe such localized inelastic interactions. Investigations of such localized inelastic scattering processes are more difficult at the lower electron energies typical of XPD because of the delocalized nature of the dominating inelastic processes (e.g. plasmons and valence-electron excitations) and the correspondingly short inelastic mean free path.

\begin{figure}
\centering
  \includegraphics[width=12cm]{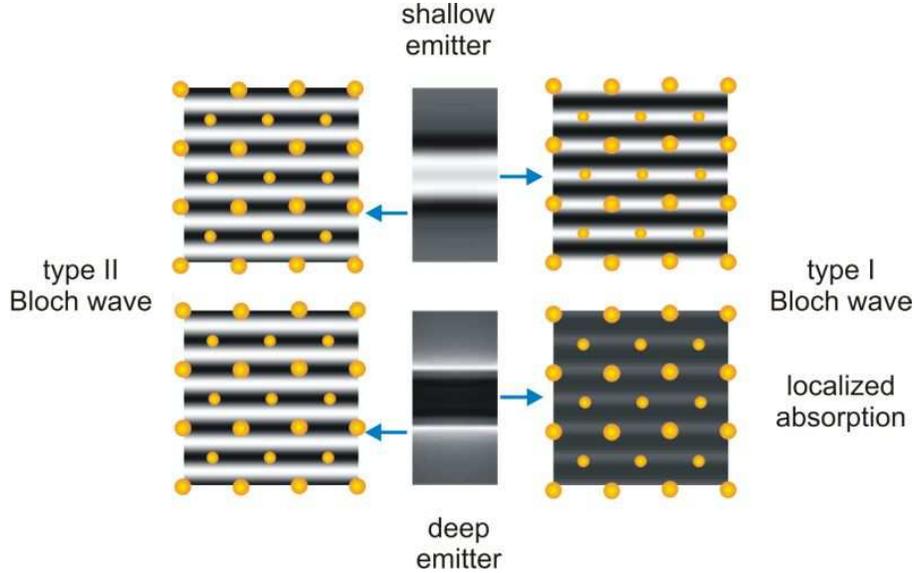}
  \caption{Contrast reversal of a Kikuchi band with increasing thickness of a photoelectron emitter.
The type I Bloch wave is localized at the atomic planes and is also
more strongly absorbed. This means that beyond a certain thickness, the type I wave is almost completely absorbed and photoelectrons do not reach the surface. Because the type I wave dominates in the middle of the band, there will me a minimum of intensity for a deep source.}
\label{reversal}
\end{figure} 

\begin{figure}
\centering
  \includegraphics[width=14cm]{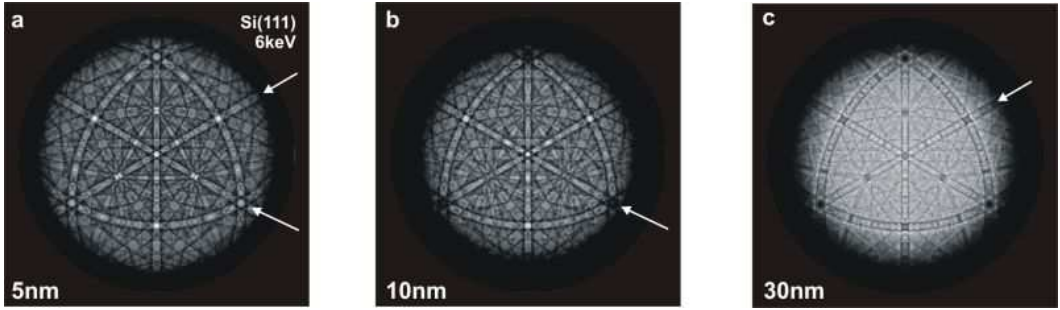}
  \caption{Contrast reversal of Kikuchi bands in photoelectron diffraction patterns from Si(111) at 6\,keV with increasing depth of emitters at Si positions starting at thicknesses 5\,nm, 10\,nm and 30\,nm. Note the reversal of contrast in the indicated features when going from 5nm to 10nm and 30nm, respectively. The inelastic mean free path was taken to be 5nm. If only isotropic absorption is considered, no contrast reversal is occurring (the patterns then are similar to Figure \ref{expsimsi}).}
\label{reversalSi}
\end{figure}

At high electron energies, we can use the crystal as an interferometric beam splitter which, in the vicinity of a Bragg reflection, distributes electrons in a very well defined way either between or onto the atomic planes. We have seen above in the discussion of the intensity distribution within a Kikuchi band, that in the middle of the band, the probability amplitude of the outgoing electrons is maximized at the atomic planes (Fig. \ref{reversal}, type I Bloch wave). This in turn also means that electrons in this wave field
will experience an increase in localized inelastic processes, like e.g. phonon losses or core excitations. In contrast, for angles slightly larger than the Bragg angle, the electron wave field is maximized between the atomic planes (Fig. \ref{reversal}, type II Bloch wave) and thus the electrons moving in this type of field can escape from the crystal with much less of these localized inelastic collisions.
If the photoemitters are distributed equally in all thicknesses of the sample, the emitters near the surface will dominate the diffraction pattern. For these low thicknesses, the difference in absorption of the two wave fields can be neglected and our explanation of section \ref{sec:blochdenXPD} is appropriate. However, if we had emitters in a buried layer, we should be able to sense the localized absorption by a contrast reversal of a Kikuchi band: with increasing thickness of the film above the emitters, the maximum of intensity in the middle of the band turns into a minimum because electrons in this wave field will be preferentially absorbed. This process will start at the larger emission angles, because these electrons have to travel the longest distance in the crystal. We illustrate this effect by calculations for Si(111) at 6\,keV in Figure \ref{reversalSi}, placing the emitters in starting depths of 5nm, 10nm, and 30nm. The Figure \ref{reversalSi} clearly shows
how the contrast is reversing with increasing thickness. A similar effect is known
in transmission electron microscopy and is termed "anomalous absorption". For the description of the localized inelastic scattering we have used the parameters of \citeauthor{birdking90} \cite{birdking90} who use an Einstein model to determine the higher order Fourier components  of the imaginary part of the potential which are mainly due to thermal diffuse scattering.
An experimental study of this contrast reversal process should thus give insight into the localization of inelastic scattering in photoemission spectroscopy.

\paragraph{Spin-dependent effects and magnetic dichroism} 

In complete analogy to the lower energy case, in HXPD it should also be possible to observe diffraction effects that depend on the
photoelectron spin \cite{vanhove1996jesrp} and magnetic dichroism. 
Although the difference in purely exchange scattering of spin-up and spin-down electrons is expected to decrease as energy is increased \cite{sinkovic1985prb}, the spin-orbit effects on scattering will increase.  Beyond this, the spin-orbit splitting of energy levels, coupled with additional energy splitting due to multiplet effects should produce  magnetic circular dichroism (MCD).
MCD in combination with HXPD should be useful for the analysis of the bulk magnetism of complex new materials.

As experimental confirmation that such effects should exist in hard x-ray photoelectron diffraction, we note that MCD effects of greater than 15\% have been seen recently in Fe 2p emission from Fe$_3$O$_4$ at 7.9\,keV excitation energy by 
\citeauthor{ueda2008apex} \cite{ueda2008apex}.  A simple one-electron picture of such effects in transition-metal 2p emission permits relating the MCD intensity to the matrix elements and phase shifts associated with the $l=\pm 1$ (s or d) outgoing wave components \cite{menchero1998prb}.
Adding into our dynamical theory a more correct description of these matrix elements should thus permit describing such magnetism-related effects and their dependence on angle.  A precise treatment of such effects would however also have to take account of the precise \mbox{L = (l,m)} states involved, so as to also include the non-magnetic circular dichroism effects associated with strong forward scattering \cite{daimon1993jjap}, although these are expected to become much smaller at higher energies due to their inverse dependence on wave vector and the distance of a given scatterer from the emitter.  Cluster calculations might in fact be a more direct way of modelling MCD in HXPD if the energies are not too high. 

\section{Comparison to Cluster Model Calculations}

\begin{figure}
\centering
  \includegraphics[width=14cm]{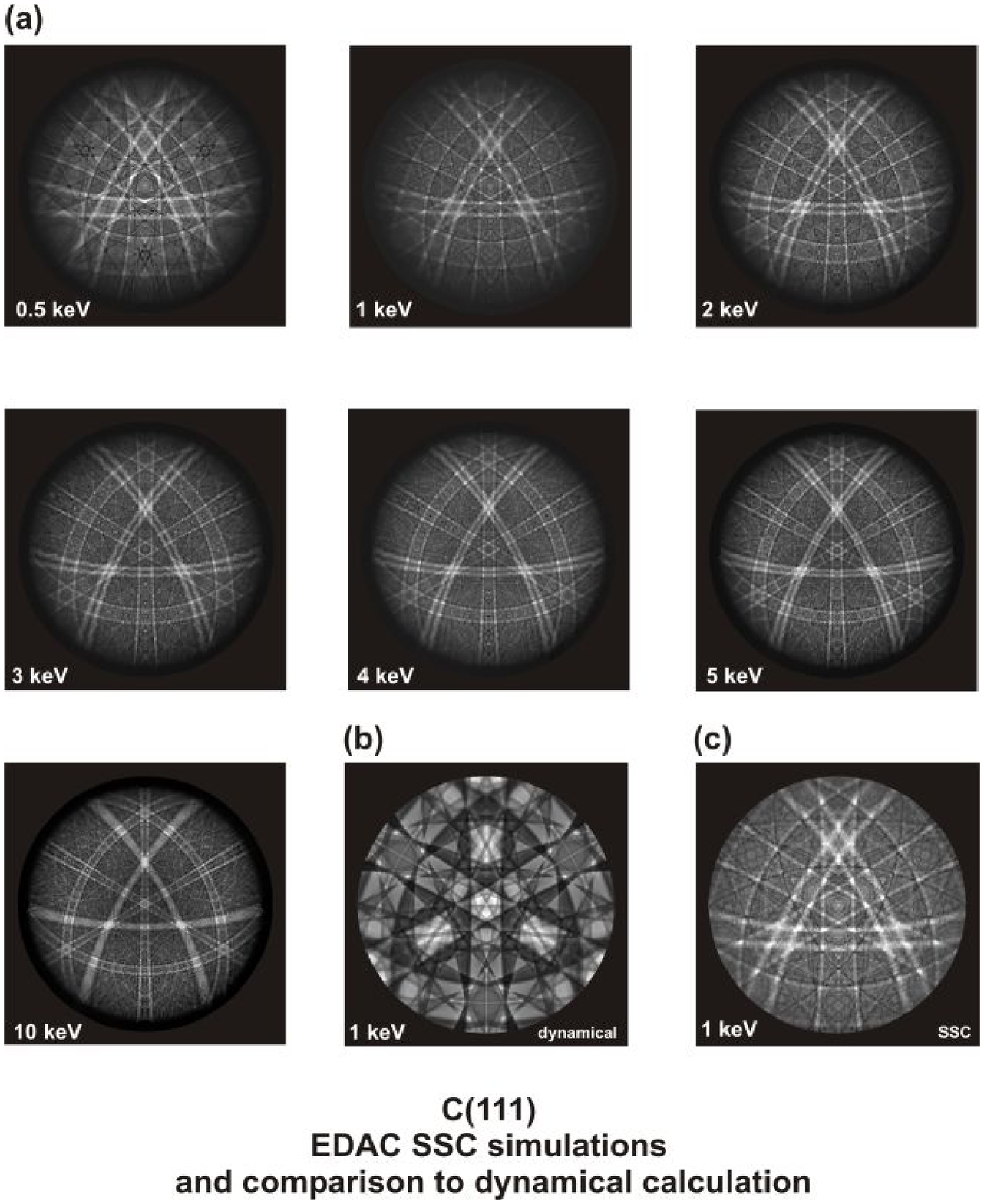}
  \caption{(a) Photoelectron diffraction patterns from diamond C(111) at the kinetic energies indicated and simulated using a single-scattering cluster (SSC) approach \cite{abajo2001prb} instead of the dynamical approach used in all prior figures.  The cluster was hemispherical and contained about 13,000 atoms. Emitters in all layers were included. (b) and (c) Comparison of the central portions of the HXPD patterns at 1\,keV emission energy from dynamical theory and SSC theory, respectively.}
\label{sscsim}
\end{figure} 

In Fig. \ref{sscsim}, we now show calculations for s-wave emission from C(111) based on a cluster model whose methodology is described in detail elsewhere \cite{abajo2001prb}. These calculations were carried out in single scattering (SS) and for emission from emitters in a cluster containing about 13,000 atoms and extending to about 30\,\AA \, below the surface.  The Kikuchi bands are clearly seen in all of these plots, which can be compared directly to the first 7 panels in Fig. 4, except that the dynamical calculations in Fig. 4 include multiple-scattering effects.  Clearly, the main trend of a narrowing of the Kikuchi bands with increasing energy is reproduced by the cluster calculation. The main difference is seen in the middle of the Kikuchi bands, for which the SS cluster (SSC) calculations exhibit decreased intensity. Even in the forward-scattering directions, which usually show enhancements of intensity, we see a reduction of intensity, in disagreement with dynamical theory and experiment at 1\,keV.
It is in fact not surprising that the SSC patterns exhibit more purely Bragg-reflection-like Kikuchi bands with less filling in between the intensity maxima, as that is exactly what in implicit in the theory.  However, even for this too-simplified model, we note that a closer inspection of the weaker fine structure in the SSC patterns shows some similarity to the dynamical patterns in Fig. 2.  In Figs. 13 (b) and (c) we compare the central portion of the patterns for 1\,keV, an energy for which we know that dynamical theory agrees with experiment (cf. Fig. 2).  Here we in particular see that those maxima which result from features at  the Bragg angle from a given set of planes generally agree between the two theoretical approaches.  However, the more subtle fine structure is definitely not correctly predicted in the SSC calculations.

Adding in multiple scattering to the all-layer cluster calculations, which becomes an even more time-consuming effort, might be expected to ultimately yield results  closer to the dynamical calculation, which implicitly includes the higher orders of scattering via the basic assumptions of the Bloch wave dynamical diffraction theory.

In summary, although the cluster approach is not the most efficient one for calculation HXPD from multilayer ordered crystals, it is able to reproduce the basic Kikuchi band features, and should still be useful in situations with less long-range order and/or for emission from structures very near the surface.

\section{Summary}
We have presented a long-range order dynamical scattering approach for simulating photoelectron diffraction patterns 
at high kinetic energies. The approach is based on exploiting the quasi
three-dimensional translational symmetry of the scattering potential
that the photoelectron experiences at high energies. The predicted patterns are dominated
by Kikuchi bands which reflect the crystallographic information, and we have verified that even at energies as low as 1\,keV, the agreement with experiment is excellent. We have discussed the systematics of such hard x-ray photoelectron diffraction (HXPD) patterns for energies up to 20\,keV and considered as well several 
possible applications, e.g. concerning tetragonal lattice distortions and impurity site determinations. These applications should extend the structural analysis power of XPD, as well as making it a much more generally useful tool for studying truly bulk properties of complex new materials.

\section*{Acknowledgements} C.S.F. acknowledges support from the Director, Office of Science, Office of Basic Energy Sciences, Materials Science and Engineering Division, U.S. Department of Energy under Contract No. DE-AC03-76SF00098, the Alexander von Humboldt Foundation, the Helmholtz Association, the J\"ulich Research Center, and the University of Hamburg during part of this work.


\end{document}